\documentstyle[prl,aps,epsf,floats]{revtex}

\begin{document}
\draft

\twocolumn[\hsize\textwidth\columnwidth\hsize\csname @twocolumnfalse\endcsname
\title{Current Rectification by Molecules with Asymmetric Tunneling Barriers}

\author{P.E. Kornilovitch, A.M. Bratkovsky, and 
R. Stanley  Williams}
 
\address{
Hewlett-Packard Laboratories, 1501 Page Mill Road, Palo Alto, 
California 94304
}

\date{\today}
\maketitle

\begin{abstract}

A simple experimentally accessible realization of current rectification 
by molecules (molecular films) bridging metal electrodes is described.  
It is based on the spatial asymmetry of the molecule and requires only 
one resonant conducting molecular level ($\pi$ orbital).  The rectification,
which is due to asymmetric coupling of the level to the electrodes 
by tunnel barriers, is largely independent of the work function difference 
between the two electrodes.  Results of extensive numerical studies of 
the family of suggested molecular rectifiers
HS-(CH$_2$)$_m$-C$_6$H$_4$-(CH$_2$)$_n$-SH are presented.  The highest 
rectification ratio $\sim 500$ is achieved at $m = 2$ and $n = 10$.   

\end{abstract}

\narrowtext
\vskip2pc]

\section{Introduction}
\label{sec:one}

The concept of a molecular rectifier started the field of molecular
electronics (moletronics) almost thirty years ago \cite{Aviram}.  At 
present, moletronics is an explosively growing field of experimental and 
theoretical activity. The molecular rectifier (MR) is still one of the 
central objects of this research.  Even the simplest future application, 
such as molecular memory, will require high quality MR with sharp voltage 
thresholds, large current rectification ratios, small time constants, 
large breakdown voltages and so on. MR demonstrated so far 
\cite{Martin,Metzger,Zhou} constitute an impressive proof of principle 
but their electrical parameters are in fact very poor.  The situation 
prompts further investigation of molecular rectification.

Until recently, theoretical analysis of MR had been largely limited to 
the Aviram and Ratner donor-insulator-acceptor ($D-\sigma-A$) mechanism 
\cite{Aviram}.  In this design, the highest occupied molecular orbital 
(HOMO) and the lowest unoccupied molecular orbital (LUMO) are confined
to two different parts of the rectifier, $D$ and $A$ respectively.
The insulating bridge $\sigma$ prevents the orbitals from ``spilling off''
to the other part.  If such a molecule is placed between two metal
electrodes, the current-voltage characteristics of the junction 
is expected to be highly asymmetric.  At a particular voltage applied
in the positive direction, the Fermi level of the electrode on $A$
side aligns with the LUMO, and the on $D$ side aligns with the HOMO.
At this voltage, the current rises sharply because the electrons 
can now be loaded on the LUMO, then tunnel {\em inelastically} through
the $\sigma$ to the HOMO and the escape into the second electrode.
In the opposite direction, a similar process does not occur until
a much higher applied voltage.  

A different mechanism of molecular rectification was described by 
Ellenbogen and Love \cite{Ellenbogen}.  It is based on an energy 
mismatch between two conducting levels localized on
different parts of the molecule.  Within the $D-\sigma-A$ framework, 
such levels could, for instance, be the LUMOs of $D$ and $A$.  Under 
external bias, the levels shift because of the electric field.  In the 
forward direction, the conducting levels align at some voltage, and 
facilitate resonant transport of electrons between the electrodes.  
In the reverse direction, the levels move away from each other and 
the current remains small.  Hence molecular rectification.  
  
We note that both described mechanisms require {\em two} electroactive
molecular levels and a fine balance between potential drops 
inside the molecule and on the molecule-electrode interfaces.
To the best of our knowledge, neither of the mechanism has been
realized experimentally.       

Recently, an interesting case of molecular rectification was discussed by 
Krzeminski {\it et al.} \cite{Krzeminski}.  They studied theoretically current 
rectification in Langmuir-Blodgett (LB) films of $\gamma$-hexadecylquinolinium 
tricyanoquinodimethanide (C$_{16}$H$_{33}$Q-3CNQ), which was previously 
studied experimentally \cite{Martin,Metzger}.  That molecule was 
initially considered to be a possible implementation of the Aviram-Ratner
mechanism because its active part Q-3CNQ comprised a donor and an
acceptor.  (The long insulating tail C$_{16}$H$_{33}$ was added to help 
form good LB films.)  It was later realized \cite{Metzger2} that the Q-3CNQ 
is, in fact, a $D-\pi-A$ molecule and is unlikely to implement the 
Aviram-Ratner mechanism.  The analysis of Ref.\cite{Krzeminski} confirmed 
that the $\pi$ bridge does not sufficiently isolate $D$ and $A$ to keep
molecular orbitals localized on either $D$ or $A$.  Instead, the orbitals
were delocalized over the entire Q-3CNQ unit.  Krzeminski {\it et al.}
have attributed
the observed rectification to asymmetric position of the LUMO and HOMO
with respect to the Fermi levels of the metal and to ``asymmetric
profile of electrostatic potential across the system''.   

The purpose of this paper is to point out that there is, indeed, a simple 
and general 
mechanism of molecular rectification where a single electroactive 
unit is positioned asymmetrically with respect to electrodes {\em and} 
the HOMO and LUMO are positioned asymmetrically with respect to the
Fermi level.  However, this mechanism {\em does not} require such complex 
molecules as C$_{16}$H$_{33}$Q-3CNQ.  What is needed is just {\em one} 
conducting molecular level placed closer to one electrode than to the other.
Since most of the applied voltage drops on the longer insulating barrier, 
the conditions for resonant tunneling through the level are achieved at 
very different voltages for the two opposite polarities.  By changing the 
lengths of the insulating barriers, the rectification ratio can be 
systematically changed.  This mechanism can be realized by relatively simple
molecules.  For instance, the conductive level could be supplied by 
a benzene ring.  Also, the molecules can be {\em shorter}, 
produced by e.g. self-assembly instead of 
LB deposition, and therefore more conductive.  Indeed, the currents reported 
in \cite{Metzger} were of order $10^{-17}$ A/molecule, which almost
rules out any practical application of such MR.  This is obviously a
result of having the long aliphatic tail C$_{16}$H$_{33}$. There is an
experimental evidence that even much shorter alkane chains, like
C$_{12}$, are very resistive and transport there proceeds by hopping
processes rather than tunneling \cite{Boulas}.
We show below that With simpler conducting 
and shorter insulating units, MR can achieve rectification ratios 
in excess of a hundred while remaining fairly conductive.

In the foregoing sections, we discuss the present mechanism of molecular 
rectification in more detail.  We have performed a numerical analysis of 
the mechanism by calculating the current-voltage (I-V) characteristics of 
the family of prototype molecular diodes 
HS-(CH$_2$)$_m$-C$_6$H$_4$-(CH$_2$)$_n$-SH. We have found that 
rectification ratios of $>100$ are achievable with such a design.

We mention for completeness that {\em any} asymmetric 
electrode-molecule-electrode junction should in principle produce 
asymmetric currents at large enough voltages.  Such a high-voltage 
asymmetry due to unequal coupling to the electrodes \cite{Xue,Ghosh} 
or to an asymmetric central molecular unit \cite{Reichert} have been 
observed experimentally and discussed theoretically.  However, the 
current rectification ratios in these studies were of order unity,
which is clearly insufficient for practical applications.  Zhou 
{\it et al} reported rectification in a molecular monolayer of 
4-thioacetylbiphenyl \cite{Zhou}.  In that experiment, the two 
electrodes were different, with different work functions, and 
different connections to the molecules.  The experimental data were 
interpreted as the standard thermionic emission through an asymmetric 
barrier.

\section{A molecular design for current rectification}
\label{sec:two}

\begin{figure}[t]
\begin{center}
\leavevmode
\hbox{
\epsfxsize=8.6cm
\epsffile{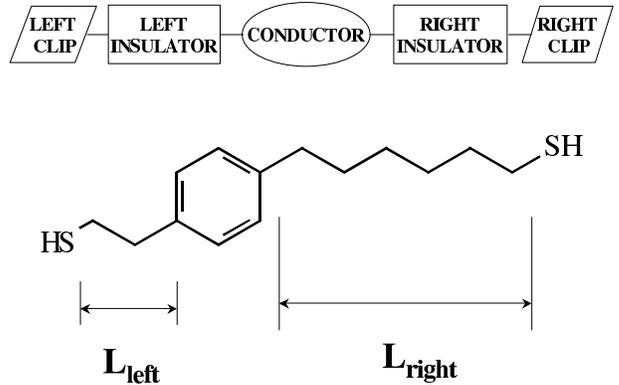}
}
\end{center}
\vspace{0.0cm}
\caption{ Top: Schematic structure of an asymmetric tunneling barrier 
molecular rectifier. Bottom: The (2,6) member of the MR 
family HS-(CH$_2$)$_m$-C$_6$H$_4$-(CH$_2$)$_n$-SH. }
\label{fig1}
\end{figure}

Our molecular rectifiers with asymmetric tunneling barriers consist 
of five structural parts, see Fig.~\ref{fig1}. The end groups provide 
contact to the two electrodes.  For instance, these may be thiols that 
could self-assemble on the surface of gold or silver, carboxyl groups 
in case the Langmuir-Blodgett technique is used, and so on.  The end 
groups may even be absent altogether if the application does not require 
them to provide better contacts.  The three inner parts of MR are the 
central conjugated group {\bf C} and two insulating barriers 
${\bf I}_{L}$ and ${\bf I}_{R}$. The purpose of such a construction is 
to provide an electronic level localized on {\bf C}, with energy not 
very different from the Fermi energies of the electrodes.  In most cases 
this level will be the lowest unoccupied molecular orbital (LUMO) of the 
molecule. Then the electron transmission probability should be resonant 
near the energy of the LUMO.

\begin{figure}[t]
\begin{center}
\leavevmode
\hbox{
\epsfxsize=8.6cm
\epsffile{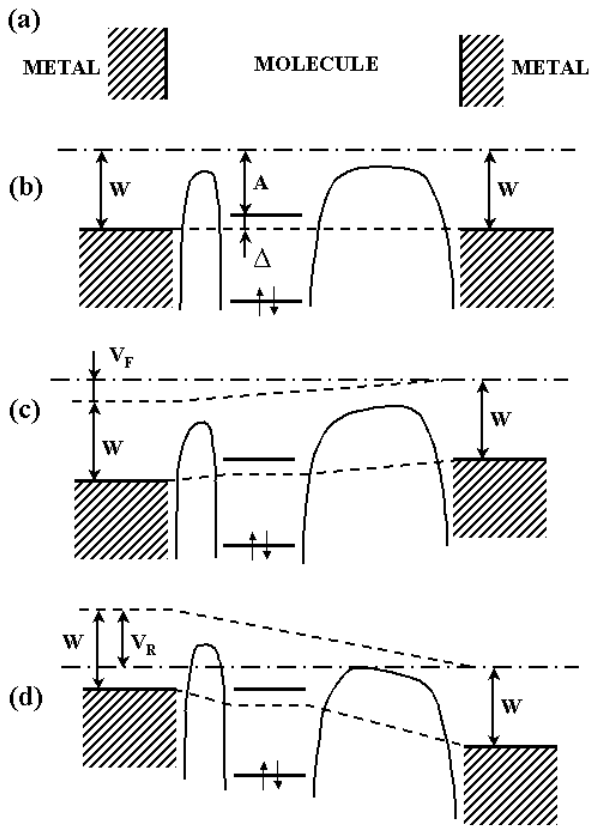}
}
\end{center}
\vspace{0.0cm}
\caption{ The basic principle of rectification by asymmetric tunnel
barriers.  $W$ is the work function of the metal, $A$ is the electron 
affinity of the molecule, $\Delta \equiv W-A$.
(b) A molecule with $A \leq W$ and with different lengths of the insulating
barriers is connected to two metallic leads. (c) Under forward bias, the
current rises when the right Fermi level aligns with the conducting
molecular level. (d) Under reverse bias, the current rises when the left
Fermi level aligns with the conducting molecular level. Since most of the
total voltage drops on the right insulation barrier, $V_R > V_F$.
}
\label{fig2}
\end{figure}

The energy diagram of MR is shown in Fig.~\ref{fig2}. The two main
parameters that determine the rectification properties of MR are the 
energy difference between the LUMO and the Fermi energy of the electrodes 
$\Delta = W - A$, and the ratio of the voltage drops on the right
and left insulating parts $\eta $. Under the assumption that the
polarizability of {\bf C} is much larger than that of the {\bf I}s, the voltage
drops on the barriers are proportional to their respective lengths 
$L_{\rm right}$ and $L_{\rm left}$, and 
$\eta \approx L_{\rm right}/L_{\rm left}$. For simplicity, 
we assume that the two electrodes are similar, or have
similar work functions. Therefore there is no contact potential difference
and the electric field on the molecule is zero at zero applied bias. The
minor complications that may arise from the contact potential difference
will be briefly discussed in Section~\ref{sec:five}. Let us consider the
operating principle of MR. We shall follow the convention that the right
electrode is always grounded and the right Fermi energy can be used as the
reference energy. Then a positive potential applied to the left electrode
shifts its electronic levels to lower absolute energies. Thus under positive
bias, the Fermi energy of the left electrode goes down. Due to the non-zero
electric field, the LUMO will be dragged down too. The energy shift of the
LUMO ${\scriptstyle\triangle } E_{\rm LUMO}$ is determined by the parameter 
$\eta $. Under our assumptions, the field across ${\bf C}$ is very small, and 
the potential drops on the two barriers, $U_{\rm left}$ and 
$U_{\rm right}$ have to sum up to the total external bias $V$, 
$U_{\rm left}+U_{\rm right}=V$. Then it is easy to show that 
\begin{equation}
{\scriptstyle\triangle } E_{{\rm LUMO}}
= - U_{{\rm right}}=\frac{\eta }{1+\eta } qV,  
\label{one}
\end{equation}
where $q$ is the elementary charge. A sharp increase in current is expected
when the LUMO lines up with the {\em right} Fermi level, that is at 
${\scriptstyle \triangle } E_{{\rm LUMO}} = \Delta $. 
The corresponding {\em forward} voltage follows from Eq.~(\ref{one}) 
\begin{equation}
V_{F} = \frac{1+\eta }{\eta }\frac{\Delta }{q}.  
\label{two}
\end{equation}
Under reverse bias, both the left Fermi level and the LUMO go up in 
energy.  The current will turn on when the {\em left} Fermi level lines up 
with the LUMO, when $\Delta = U_{{\rm left}}=V-U_{{\rm right}}$. 
This corresponds to a {\em reverse} voltage 
\begin{equation}
V_{R} = (1+\eta )\frac{\Delta }{q}.  
\label{three}
\end{equation}
Clearly, the forward and reverse voltages are different, their ratio being 
$V_{R}/V_{F}=\eta $. At a large $\eta $, the two thresholds differ, and 
there is a voltage window $V_{F}<|V|<V_{R}$ within which there is a 
substantial current in the forward direction and almost no current in 
the reverse direction. Hence, a strong rectifying effect is expected.

Different parts of MR affect different properties of the I-V
characteristic. Parameter $\Delta$ determines the overall scale of the
forward and reverse voltages. $\Delta$ itself could be systematically
changed by using different materials for the electrodes and different
conductive parts {\bf C}. The work functions of the electrodes used in
moletronics studies vary from 5.6 eV (Pt) to 4.0 eV (Si), enabling
almost continuous  
adjustment of $\Delta$. (Obviously, any changes of the electrode material 
have to be compensated by respective changes of the side groups of MR.) 
The use of different materials for the two electrodes provides further 
tuning of $\Delta $. Indeed, the contact potential changes energy of the 
LUMO by some additional amount. [Cf. Eq.~(\ref{one}) and use the contact 
potential in place of bias $V$.] The conclusion from this qualitative 
analysis is that $\Delta$ is a parameter under the control of a 
designer of molecular rectifiers. 
What absolute value of $\Delta$, large or small, is optimal from the 
electronic viewpoint is not clear a priori.  Increasing $\Delta$ widens 
the rectification window and improves the stability of the device. At the 
same time, larger $\Delta$s imply higher operating voltages, power 
dissipation and other unwelcome consequences.

The shortest of the two insulating barriers controls the width of the
conducting level and consequently the sharpness of the current increase in
the vicinity of $V_{F}$ and $V_{R}$. Indeed, the transmission probability 
$T(E)$ through a molecular level, which defines the 
conductance of the molecule $G(E)\propto T(E)$ according to standard
Landauer formalism\cite{Landauer}, is well described by the Breit-Wigner
formula 
\begin{equation}
T(E)=\frac{\Gamma _{L}\Gamma _{R}}{(E-E_{{\rm MO}})^{2}+\frac{1}{4}
(\Gamma_{L}+\Gamma _{R})^{2}},  
\label{eq:TE}
\end{equation}
where $E_{\rm MO}$ is the energy of the molecular orbital available for
resonant transmission (LUMO in the present case), and $\Gamma _{L,R}$ are the
partial widths associated with coupling to the left and right electrodes,
respectively, see analysis in \cite{KB01}. The ``golden rule'' estimate
gives $\Gamma _{L,R}\approx t_{L,R}^{2}/D_{L,R}$, $t_{L,R}$ being the
effective coupling matrix element and $D_{L,R}$ the electron bandwidth 
in the electrodes.  We see that a steep rise in current occurs when one 
of the electrode Fermi levels rises to line up with $E_{\rm MO}$, which 
is exactly what is required for large rectification ratios.  If the 
{\bf C} component of the molecule is too close to a metal, the conducting 
level is so broad that the transmission probability will be substantial 
at all energies. Indeed, $\Gamma _{i} \propto e^{-2\kappa L_{i}},$ where 
$\kappa \propto (E_{\rm barrier}-E_F)^{1/2}$ is the tunneling 
attenuation coefficient.  Here $E_{\rm barrier}$ is the energy of 
of the conducting level of the insulating {\em barrier}, which lies
higher that the LUMO of the conducting unit {\bf C}.  For alkane chains
on gold, $(E_{\rm barrier}-E_F) \approx 4.8$ eV \cite{Boulas}.  
Thus the width of the level is exponentially sensitive to the distance 
from the electrode. If the level is substantially broadened, there will 
be little difference between the currents in the forward and reverse 
directions.  Sufficient insulation of the conducting molecular 
orbital from {\em both} electrodes is an essential feature of the 
present rectification mechanism.

Finally, the longest of the two barriers controls the anisotropy of the
I-V characteristic and the overall amplitude of the current. 
As was discussed above, $V_{R}/V_{F}\approx L_{\rm right}/L_{\rm left}$.
Therefore the longer the second barrier the larger the $V_{R}/V_{F}$ ratio
and the better the diode. At the same time, the current goes down
exponentially with the barrier length. Indeed, the current is 
\begin{eqnarray}
I & = &\frac{2q}{h}\int dE\left[ f\left( E-\frac{qV}{2}\right) -
f\left( E+\frac{qV}{2}\right) \right] T(E)  \nonumber \\
& = &\frac{2q}{\hbar }\frac{\Gamma _{L}\Gamma _{R}}{\Gamma _{L}+\Gamma _{R}}
\approx \frac{2q}{\hbar } \Gamma_{R} \propto e^{-2\kappa L_{>}},
\label{eq:Ires}
\end{eqnarray}
when the resonance falls into the ``window'' between the lowest and the
highest Fermi levels in the leads. The current falls off exponentially with
the thickness of the {\em thicker} of the insulating barriers $L_{>}$.
Therefore, we are facing a typical trade-off problem but in its worst 
form. The asymmetry of MR improves linearly but the resistance grows 
exponentially with the length of the barrier. Nonetheless, we found that 
substantial rectification is achieved at the current levels that make 
electronic applications of MR practical. This is detailed in the 
next sections.

\section{Calculation details}
\label{sec:three}

In order to test the proposed mechanism of molecular rectification, we
calculated the I-V characteristics of a prototype family of molecular 
diodes HS-(CH$_2$)$_n$-C$_6$H$_4$-(CH$_2$)$_m$-SH sandwiched between two 
gold electrodes. The five functional parts of MR are constructed from 
the standard chemical groups used in moletronic studies. The end thiols
-SH chemically attach the molecules to gold, the middle benzene ring
provides a conducting level (LUMO) at energy $E_{\rm LUMO} = -3.5$
eV (with respect to vacuum), and the insulating barriers are made of 
saturated hydrocarbon units -CH$_2$-. Individual members of the family 
are parameterized by the two integers $(n,m)$ which are the respective 
numbers of the -CH$_2$- groups on both sides of the molecules.

\begin{figure}[t]
\begin{center}
\leavevmode
\hbox{
\epsfxsize=8.0cm
\epsffile{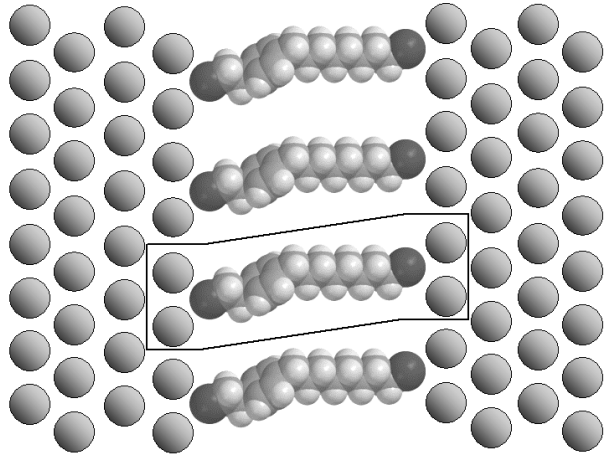}
}
\end{center}
\vspace{0.0cm}
\caption{A monolayer of molecular diodes between two semi-infinite fcc
electrodes. Only one layer of electrode atoms is shown. The solid line
encloses the atoms included in one molecular complex. 
}
\label{fig3}
\end{figure}

The electrode-MR-electrode junction is shown in Fig.~\ref{fig3}. The
equilibrium structure of isolated molecules was obtained through total
energy minimization with the density-functional program Spartan 
\cite{Spartan}. An important factor affecting the electrical properties 
of the junction is the local geometry of the molecule-electrode contact. 
It is generally accepted that, upon self-assembly on gold, thiols lose the 
end hydrogens and then bind directly to the gold atoms. The preferential 
binding position of sulfur atoms on the Au(111) surface is at the apex of the
pyramid with a triangular base of gold atoms ({\em hollow} position).
Also possible is the {\em top} position, where the sulfur is positioned
directly above one of the gold atoms \cite{Sellers}. In this paper, we assume
the hollow binding position of the sulfur. 
In the hollow position, the end S states strongly hybridize with
states on gold atoms, and the resulting current is less dependent on
exact molecule-contact geometry in comparison to the top position\cite{BK02}.
Starting with the equilibrium
molecular structure we remove the end hydrogens and replace them with two
clusters of three gold atoms. The clusters form $\sigma$-bonds to sulfurs
instead of the lost hydrogens. The primary molecule axis is roughly 
perpendicular to the planes of the gold triangles. The resulting molecular 
complexes (three gold atoms -- dithiolate -- three gold atoms) are
organized in a periodic two-dimensional film which is placed between two
(111) surfaces of semi-infinite gold electrodes. The film is commensurate
with the Au(111) surface but its primitive surface cell is four times 
larger.

In order to obtain the I-V characteristic of the device we use
the Landauer formulation of quantum transport \cite{Landauer} together with
a semi-empirical tight-binding parameterization of the Hamiltonian matrices
for the molecule and the electrodes. The off-diagonal matrix elements of the
molecule are taken from Harrison \cite{Harrison}. They all are Slater-Koster
linear combinations \cite{Slater} of the four basic elements 
$W_{\alpha \beta \gamma }$ 
\begin{equation}
W_{\alpha \beta \gamma }(i-j)=7.62\frac{\eta _{\alpha \beta \gamma }}
{d_{ij}^{2}}~{\rm eV},  
\label{four}
\end{equation}
where $d_{ij}$ is the distance between atoms $i$ and $j$, and 
$\eta_{ss\sigma }=-1.40$, $\eta _{sp\sigma }=1.84$ 
$\eta _{pp\sigma }=3.24$, and $\eta _{pp\pi }=-0.81$. 
The diagonal matrix elements, that is on-site energies 
$\varepsilon _{i}$ of {\em core} atomic orbitals, are also adopted
from \cite{Harrison}, while the on-site energies of valence orbitals are
calculated from atomic affinities $A_{i}$ and ionization potentials 
$I_{i}$ as discussed in \cite{BK02}
\begin{equation}
\varepsilon _{i} = -\frac{I_{i}+A_{i}}{2}.  
\label{five}
\end{equation}
The affinity and ionization potential determine the atomic Hubbard 
parameter $U_{i}$ as well: 
\begin{equation}
U_{i} = I_{i} - A_{i} .  
\label{six}
\end{equation}
Table~\ref{table1} lists the molecular tight-binding parameters used in 
this paper.

\begin{table}[t]
\begin{tabular}{|c||c|c|c|c|}
& $\varepsilon_s$ (eV) & $\varepsilon_p$ (eV) & U (eV) & Number of electrons \\ 
\hline\hline
C & -6.91 & -17.52 & 8.7 & 4 \\ \hline
H & -7.85 & -- & 11.5 & 1 \\ \hline
S & -20.80 & -6.47 & 7.8 & 6 \\ \hline
Au & -5.88 & -- & 6.7 & 1
\end{tabular}
\vspace{0.5cm}
\caption{ Tight-binding parameters of atoms used in transport calculations. }
\label{table1}
\end{table}

Our parameterization for the electrodes is based on the parameter sets of
Papaconstantopoulos \cite{Papa}. Although our calculational procedure
enables us to treat the leads with the full set of $s$, $p$, and $d$
orbitals \cite{KB01,Sanv99}, for the purposes of this paper the details of
the electrode electronic structure are not important. We therefore choose to
work with the $s$ component of the band structure only.  The $W_{ss\sigma }$ 
parameter of Ref.\cite{Papa} for gold is $-0.909$ eV.  We found, however, 
that this value results in too narrow a band with the high-energy edge being
not very far from the conducting LUMO of MR. In order to separate molecular
rectification from the specific effects associated with gaps in the leads'
density of states, we also used the double value, $W_{ss\sigma }=-1.818$ eV.
The same value of $W_{ss\sigma }$ is used for matrix elements inside the
electrode and between the electrode surface and atomic triangles that are
part of the molecules. A comparison of the results obtained using the
two values of $W_{ss\sigma }$ will be given in Section~\ref{sec:four}. 
The second important parameter is the Fermi energy of the electrodes 
(that is minus the work function of the lead material). The effectiveness 
of our rectification mechanism depends critically on $E_{F}$. Therefore we 
choose to study a {\em series} of $E_{F}$ which imitates the effect of 
using different electrodes. In particular, we studied the six values 
$E_{F}=-4.0$, $-4.2$, $-4.5$, $-5.0$, $-5.2$, and $-5.5$ eV.

Calculation of current is performed utilizing a multi-step procedure.
First, an isolated molecular complex (three Au atoms -- dithiolate -- three
Au atoms) is treated as follows. The molecular Hamiltonian $H_{{\rm mol}}$
is constructed from the parameters described above. Then the molecular wave
functions and their energies are found by diagonalizing the matrix 
$E-\hat{H}_{{\rm mol}}$. From the wave functions, the average number of 
electrons on each atom is calculated: 
\begin{equation}
q_{i} = 2\sum_{\alpha _{i}}\sum_{n}\left| \psi _{n\alpha _{i}}\right| ^{2},
\label{seven}
\end{equation}
where the index $n$ numbers the molecular orbitals and $\alpha _{i}$ the 
atomic orbitals that belong to atom $i$. After that the diagonal matrix 
elements of $H_{{\rm mol}}$ (the onsite energies $\varepsilon _{i})$ are 
recalculated as 
\begin{equation}
\varepsilon _{i}^{\prime }=\varepsilon _{i}+U_{i}(q_{i}-Z_{i}),
\label{eight}
\end{equation}
where $Z_{i}$ is the atomic charge of the $i$-th atom, which in turn 
changes the charges $q_{i}$ on the atomic sites. Then the procedure is 
repeated until the charges $q_{i}$ converge. The converged charges 
define the final position of the molecular levels and the molecular 
Hamiltonian to be used in the transport calculations.

In the second step, the semi-infinite electrodes are solved. In our approach,
there are two global quantum numbers, total energy $E$ and momentum parallel
to the surface ${\bf k}_{\parallel }$. Fixing ${\bf k}_{\parallel }$
converts the semi-infinite three-dimensional problem into a semi-infinite
one-dimensional problem with a finite dimensional basis. Notice, however,
that the Hamiltonian of the one-dimensional wire explicitly depends on 
${\bf k}_{\parallel }$. Thus transport problems at different 
${\bf k}_{\parallel }$ are not equivalent. Moreover, the dependence on 
this quantum number could be substantial. It is important therefore to 
consider a grid of ${\bf k}_{\parallel }$ and the grid should be as 
dense as possible. Then we follow the procedure of Ref.~\cite{Sanv99}, 
modified for the presence of an arbitrary oriented surface. For each $E$ and 
${\bf k}_{\parallel }$, we solve the {\em channel} problem, i.e. find all 
the Bloch vectors with both real and complex $k_{z}$ vectors. The former 
correspond to open conducting channels while the latter to evanescent 
channels. The wave functions and $k_{z}$ values of the channels are used 
to construct the surface Green's functions of the electrodes. Note that 
no energy or momentum integration is required with this method.

On the next step, the molecule is eliminated from the picture by first
solving the Schr\"{o}dinger equation for the molecular wave function and
then substituting the solution into the Schr\"{o}dinger equation for the
surface atoms of the leads. As a result of such a procedure, a matrix
operator $\hat{V}$ appears that directly couples states on the left wire
with the states on the right wire. $\hat{V}$ is essentially the inverse of 
$E-\hat{H}_{{\rm mol}}$ convoluted with the matrices that describe the
molecule-electrode coupling. Therefore, it contains all the  information 
about the molecule. In particular, $\hat{V}$ has poles at the energies of 
the molecular orbitals. Knowing $\hat{V}$ and the Green's functions of the 
free electrodes, the full electrode Green's function is found from the Dyson 
equation \cite{Sanv99}
\begin{equation}
\hat{G}=(\hat{G}_{0}^{-1}-V)^{-1},  
\label{nine}
\end{equation}
where $\hat{G}_{0}$ is the block-diagonal matrix in which the upper left 
corner is the free surface Green's function of the left wire and the bottom 
right corner is that of the right wire. Note that the size of all the 
matrices involved in Eq.~(\ref{nine}) is equal to the combined dimension 
of the left and right electrode surfaces. In the present study, each 
surface unit cell has four states, therefore all the matrices are 
$(8\times 8)$.

Since the Green's function solves Schr\"odinger's equation for
non-coincident spatial arguments, the transmission coefficients 
$t_{nn^{\prime }}(E,{\bf k}_{\parallel })$ between the open channels
can be found by multiplying $\hat{G}$ by certain projector vectors 
$\hat{P}$. Each projection recovers an incident wave in one of the open 
channels in a particular electrode. The components of the vector 
$\hat{G}\hat{P}$ determine the transmission and reflection amplitudes 
in all the open channels of both electrodes \cite{Sanv99}.  The 
contribution to current is found by squaring the transmission 
coefficients and summing over the open channels of the receiving wire. 
The total current is obtained by summing over all the incident channels 
available, then over ${\bf k}_{\parallel }$, and integrating over 
energy $E$: 
\begin{eqnarray}
I(V)&=&\frac{2q}{h}\sum_{{\bf k}_{\parallel}}
\int_{E_{F}-qV/2}^{E_{F}+qV/2}dE T(E)\\
T(E)&=&\sum_{nn^{\prime }}
\left| t_{nn^{\prime }}(E,{\bf k}_{\parallel })\right| ^{2}.  
\label{ten}
\end{eqnarray}
In the present study, we have used a grid of 64 ${\bf k}_{\parallel }$
points uniformly distributed over the two dimensional surface Brillouin 
zone.

An important and technically difficult problem is to account for the additional
charge that the molecule acquires due to the mismatch of its equilibrium
chemical potential with the Fermi energy of the electrodes. All the
molecular components of the scattering wave functions could be found from
the same Green's function (\ref{nine}). Squaring them and integrating over
energy, one finds new charges on the molecule, much like Eq.~(\ref{seven}).
Then the new on-site molecular energies are computed according to 
Eq.~(\ref{eight}). Thus in principle, the entire calculational procedure 
described above (except the final calculation of current) has to be 
repeated many times until the charges converge. Moreover, this self-consistent
procedure has to be performed anew for every value of the external 
bias voltage $V$, i.e. out of equilibrium. The charges in this
non-equilibrium case
can be calculated with the use of the procedure described in \cite{BK02}.

The additional charge transfer is most important for 
strong coupling between the molecule and the wires. In this case, the 
molecular levels broaden significantly so that their tails can accommodate 
significant additional charge. Luckily, our rectification mechanism works 
best for narrow molecular resonances. As long as the Fermi energy remains 
below the conducting molecular level, the additional charge is negligible. 
For this reason, we do not perform the self-consistent calculation of 
charge in this paper. What we do take into account is the linear shift 
of the on-site atomic energies due to the external electric field due 
to the bias voltage.

\section{Results}
\label{sec:four}

\begin{figure}[t]
\begin{center}
\leavevmode
\hbox{
\epsfxsize=8.6cm
\epsffile{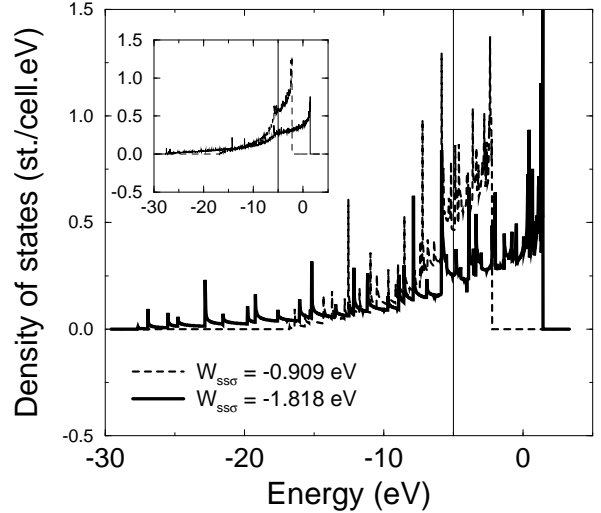}
}
\end{center}
\vspace{0.0cm}
\caption{ The bulk electrode density of states for the two values of 
$W_{ss\protect\sigma }$. The data are collected from a $(8\times 8)$ grid of 
${\bf k}_{\parallel }$ points, the one-dimensional square-root 
singularities being visible. The vertical solid line indicates 
the position of the Fermi level. Inset: the same density of states but 
computed on a $(40\times 40)$ grid of ${\bf k}_{\parallel }$ points. }
\label{fig4}
\end{figure}

We begin by presenting our numerical results for the electrode density of
states $N(E)$ (DOS), see Fig.~\ref{fig4}. Clearly visible are the square root
singularities, which are characteristic of one-dimensional conductors. Recall
that fixing ${\bf k}_{\parallel}$ renders the wire one-dimensional and the
three-dimensionality is restored upon summation over the infinite number of 
${\bf k}_{\parallel}$ points. The comparison of the main panel and inset in
Fig.~\ref{fig4} indicates that this is indeed the case. Our choice of 64 
${\bf k}_{\parallel}$ points is a reasonable compromise between the accuracy
and time for the transport calculations. The DOS affects the
I-V characteristic of the diode in two ways.  First, it changes
the effective resonance width since $\Gamma \propto N(E)$. Thus, in the 
case of $W_{ss\sigma} = -1.818$ eV, the resonances are expected to
be twice as narrow as for $W_{ss\sigma} = -0.909$ eV ($N \propto D^{-1}$.)
Secondly, gaps/edges in the density of states could block elastic
tunneling completely, resulting in a strong negative differential
resistance effect.  This possibility is discussed in Subsection (iv)
below.

We investigated four major effects of the spatial and electronic structure
of the junction on its I-V characteristic.

\begin{figure}[t]
\begin{center}
\leavevmode
\hbox{
\epsfxsize=8.6cm
\epsffile{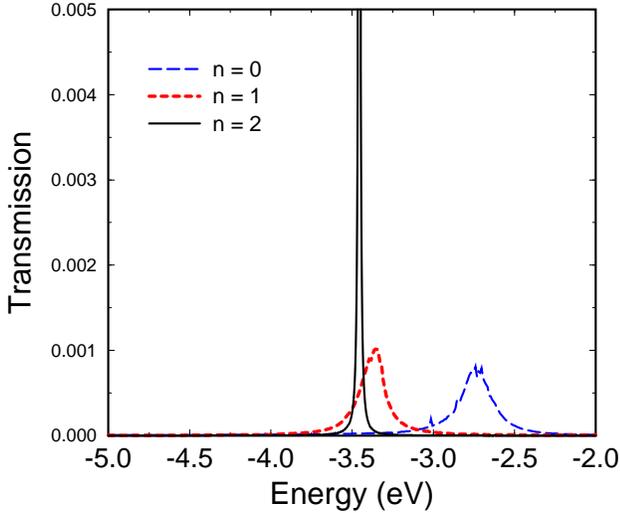}
}
\end{center}
\vspace{0.0cm}
\caption{ Transmission probability through molecular diodes 
Au$_3$-S-(CH$_{2}$)$_{n}$-C$_{6}$H$_{4}$-(CH$_{2}$)$_{6}$-S-Au$_3$, 
$n=0,1,2$.  The peak corresponds to transmission through the LUMO 
located on the benzene unit. }
\label{fig5}
\end{figure}

{\em (i) Effect of the width of the molecular level on the rectification
property of the junction.} Within our mechanism, good rectification requires
narrow molecular resonances. If the levels are narrow, the current rises
steeply upon reaching the threshold voltage. In other words, there is very
little current before the resonance is reached but some finite current
after, which can produce a high rectification ratio. In the
case of a broad resonance, the Lorentzian tails of the transmission function
provide a sizeable current even before the nominal threshold voltage is 
reached. The rectification ratio should be small in this case. The width
of the resonance is controlled by the thickness of the {\em shorter}
insulating barrier. In Fig.~\ref{fig5} we compare the transmission function
for the three molecular rectifiers which have no, one, and two insulating 
-CH$_{2}$- groups on the short side. One can see that in the first two cases 
the resonance has a sizeable width of 0.2 eV (full width at half maximum), 
which implies strong coupling to the electrode. This is more or less obvious 
for $n=0$, when the sulfur atom is directly attached to the benzene. The lone
electron pair of the sulfur overlaps with the $\pi $ electrons of the ring,
resulting in a molecular orbital distributed almost evenly over the ring and
the sulfur. Since the sulfur is directly coupled to the electrode
(albeit through an intermediate triangle), the molecular orbital is
significantly broadened. In the $n=1$ rectifier, the sulfur is separated from
the ring by a single insulating -CH$_{2}$- group. However, because of the 
$sp^{3}$ hybridization of carbon, the sulfur atom is out the plane of the 
ring. Then the separation between the sulfur and the ring is not large 
enough to prevent elongated $p$-orbitals of sulfur and carbon from direct 
overlapping.  As a result, the electronic level is still significantly 
broadened. The situation changes radically for $n=2$. Two insulating 
groups move the sulfur away from the ring by 4.3 \AA, so that direct 
overlap between the sulfur and the ring wave function becomes small. 
(From this point, insertion of new groups results in an exponential 
decrease of overlap.) According to our calculations, for $n=2$ the width 
is just $\sim 10$ meV, see Fig.~\ref{fig5}.

The numerical results illustrate the general formula for transmission
probability $T(E)$, Eq.~(\ref{eq:TE}). For the present molecular diodes, 
one partial level width is much larger than the other, say 
$\Gamma _{L}\gg \Gamma _{R}$. Then transmission at the resonance is 
$T(E=E_{{\rm MO}})=4\Gamma _{R}/\Gamma _{L}$. Thus, as $\Gamma _{L}$ 
decreases due to better insulation, the resonance gets narrow but higher. 
For the $n=2$ diode, the resonance is narrow enough to be comparable with 
thermal, disorder, and other types of broadening present in the system. 
Further increase of the short barrier is unnecessary, it will result only in 
reducing the molecular asymmetry and spoiling the rectification property. 
Our conclusion from this analysis is that {\em two} -CH$_{2}$- 
{\em groups on the short side is the optimal choice} for rectification. 
One should add that this rule has been derived for thiol-terminated 
molecules that are self-assembled on gold or other noble metal. For other 
types of contact, this may change. For instance, when an LB film is 
deposited on a metallic surface, a larger (Van-der-Waals-like) gap 
may exist between the film and the metal. Such a gap will serve as an 
additional insulator, which may reduce the optimal length of insulating 
material on the molecule itself.

\begin{figure}[t]
\begin{center}
\leavevmode
\hbox{
\epsfxsize=8.6cm
\epsffile{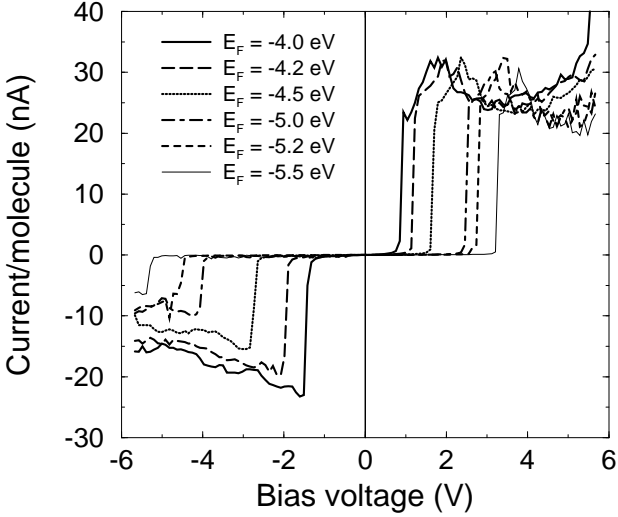}
}
\end{center}
\vspace{0.0cm}
\caption{ The effect of the metal work function on the I-V
characteristic of the molecular diode 
-S-(CH$_{2}$)$_{2}$-C$_{6}$H$_{4}$-(CH$_{2}$)$_{6}$-S-. }
\label{fig6}
\end{figure}

{\em (ii) Effect of the metal work function on the forward and reverse
voltages.} According to Eqs.~(\ref{two}) and (\ref{three}), the forward and
reverse voltages are directly proportional to the equilibrium energy
difference $\Delta$ between the conducting molecular level and the Fermi 
energy of the metal. We have studied this effect by varying the electrode 
Fermi energies with respect to the molecular orbitals. The results are 
presented in Fig.~\ref{fig6} and in Table~\ref{table2}. The I-V
characteristics all have similar shape but their forward and reverse
voltages systematically increase with $\Delta$, as expected. Analysis of
parameters from Table~\ref{table2} reveals that the ratios 
$V_F/\Delta \approx 1.6$, $V_R/\Delta \approx 2.6$, and 
$V_R/V_F \approx 1.6$ all remain approximately constant as functions of 
$\Delta$ for $m=2$ and $n=6$. Thus the ratio $V_R/V_F$ is independent
of electrode material  
and is indeed a characteristic of the molecule asymmetry only. However, 
the relation $V_R/V_F = \eta \approx L_{{\rm right}}/L_{{\rm left}}$ is 
satisfied only approximately. For the (2,6) molecule, presented in 
Fig~\ref{fig6}, $L_{{\rm right}}/L_{{\rm left}} = 3$ which is almost 
twice the voltage ratio. This is because the simple equations (\ref{two}) 
and (\ref{three}) do not take into account the voltage drop on the central 
conductive unit.  Clearly, some voltage always drops there, which effectively 
increases the lengths of both barriers and reduces the anisotropy. 
We discuss this issue in more detail in subsection (iii) below.

\begin{table}[b]
\begin{tabular}{|c||c|c|c|c|c|c|}
$W$ (eV) & $\Delta$ (eV) & $V_F$ (V) & $V_R$ (V) & $V_R/V_F$ &
$V_{{\rm op}}$ (V)
& $I_{+}/I_{-}$ \\ \hline\hline
4.0 & 0.55 & 0.90 & 1.44 & 1.60 & 1.13 & 56  \\ \hline
4.2 & 0.75 & 1.19 & 1.93 & 1.62 & 1.41 & 118 \\ \hline
4.5 & 1.05 & 1.66 & 2.73 & 1.64 & 2.07 & 173 \\ \hline
5.0 & 1.55 & 2.51 & 4.01 & 1.60 & 3.02 & 276 \\ \hline
5.2 & 1.75 & 2.77 & 4.50 & 1.63 & 3.68 & 305 \\ \hline
5.5 & 2.05 & 3.25 & 5.32 & 1.64 & 4.25 & 323
\end{tabular}
\vspace{0.5cm}
\caption{ Parameters of the I-V characteristics shown in 
Fig.~\ref{fig6} for $m=2$ and $n=6$. $\Delta = W - |E_{{\rm LUMO}}|$
is computed from the  
work function $W $ and the molecular orbital energy 
$E_{{\rm LUMO}} = -3.45$ eV. $V_F$ and $V_R$ are the current onset 
voltages under positive and negative bias, respectively. 
$V_{{\rm op}}$ is some {\em operating} voltage between $V_R$ and $V_F$. 
The last column represents typical current rectification ratios 
$I_{+}/I_{-} \equiv I(+V_{{\rm op}})/I(-V_{{\rm op}})$. }
\label{table2}
\end{table}

The last two columns of Table~\ref{table2} present data on current
rectification. As a measure of the latter, we choose to compare currents at
some positive and negative voltages with the same absolute value 
$V_{{\rm op}}$. Obviously, rectification is strongest if $V_{{\rm op}}$ 
is chosen between $V_F$ and $V_R$. In this case, in the positive direction 
there is already some appreciable current due to resonant tunneling 
through the molecular level. In the reverse direction, the tunneling 
is still under barrier and therefore is exponentially reduced. This is 
the essence of our rectification mechanism. The values of the operating 
voltage $V_{{\rm op}}$ shown in the table are roughly halfway between 
$V_F$ and $V_R$. (Approximately, $V_{\rm op} \approx 1.3 V_F$.)
One can see that the current ratio is a steadily 
increasing function of $\Delta$. The overall dependence is close to 
linear, with both end points being slightly off this general trend.

An interesting problem is choosing an optimal value of $\Delta$. As we have
seen, large $\Delta$s result in better rectification of current and wider
rectification voltage windows $V_R - V_F$. The latter fact is very important
because all sorts of disorder in the system will tend to shrink the working
window. Also, a particular circuit design may require the {\em operating} 
reverse voltage must be larger in absolute value than a certain voltage.  
In general, the absolute value $V_R$ has to be as large as possible. Both
these arguments favor a large $\Delta$. On the other hand, large operating
voltages imply higher power and higher electrostatic stress on
the molecules, which are, of course, undesirable. Quantitative 
understanding of this trade off problem warrants a detailed investigation
for a particular application.

\begin{figure}[t]
\begin{center}
\leavevmode
\hbox{
\epsfxsize=8.6cm
\epsffile{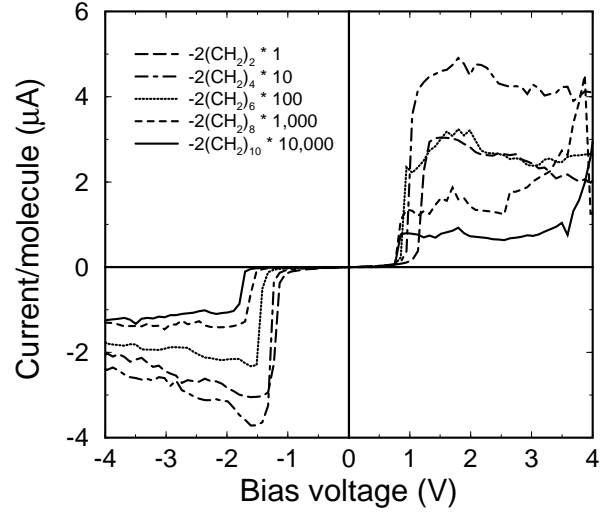}
}
\end{center}
\vspace{0.0cm}
\caption{ Dependence of the I-V characteristic on the asymmetry
of the molecular diode HS-(CH$_2$)$_2$-C$_6$H$_4$-(CH$_2$)$_n$-SH for 
work function $W = 4.0$ eV. Notice how the $V_F$ goes down while $V_R$ 
goes up, systematically shifting the I-V curve to the left from the origin. }
\label{fig7}
\end{figure}

{\em (iii) Effect of the length of the insulating barrier on the asymmetry
of the I-V characteristic and on the overall current magnitude.}  This 
effect is central to the present paper. By increasing the length of the 
{\em longer} barrier while keeping the shorter one fixed, one forces a
larger portion of the voltage to drop on the longer barrier. This implies that
the conducting threshold is reached at different bias voltages in the two
opposite directions, as illustrated in Fig.~\ref{fig2}. This design
principle provides us with the means to systematically increase the current
rectification ratio and produce better molecular diodes. In Fig.~\ref{fig7},
we show the calculated I-V characteristics of a series of molecular diodes 
$(2,n)$ for work function $W=4.0$~eV.
The electrical parameters of the I-V characteristics are summarized
in Table~\ref{table3}. The series begins with the symmetric molecule $n = 2$,
which produces a symmetric I-V characteristic with $V_R = V_F = 1.19$ V. 
With increasing $n$, $V_F$ monotonically decreases but $V_R$ increases, 
resulting in a systematic shift of the I-V characteristic with respect 
to the origin. By $n=10$, the reverse-to-forward voltage ratio reaches 
$V_R/V_F = 2.18$. The voltage ratio grows slower than $\propto n/2$,
suggested by Eqs.~(\ref{two}) and (\ref{three}). It could be described by 
the relation 
\begin{equation}
\frac{V_R}{V_F} = \frac{L_{{\rm right}} + {\scriptstyle \triangle} L} 
{L_{{\rm left}} + {\scriptstyle \triangle} L} ,  
\label{twelve}
\end{equation}
where ${\scriptstyle \triangle} L$ is a correction that takes into account
the voltage drop on the central conducting unit. Using the data of 
Table~\ref{table3}, one finds 
${\scriptstyle \triangle} L/L_{{\rm left}} = 2.5$, 4.0,
4.0, and 4.9 for $n = 4$, 6, 8, and 10, respectively. The correction does
not remain constant as a function of $n$. Therefore it is difficult
to assign to it a useful geometrical meaning.

The penultimate column of Table~\ref{table3} shows the current
rectification ratios $I_{+}/I_{-}$. It grows approximately 
proportional to $(n-2)$, reaching a value of $\sim 100$ at $n=10$. 

\begin{table}[b]
\begin{tabular}{|c||c|c|c|c|c|c|}
$n$ & $V_F$ (V) & $V_R$ (V) & $V_R/V_F$ & $V_{{\rm op}}$ (V) & $I_{+}/I_{-}$ & 
$R ({\rm M}\Omega$ \\ \hline\hline
2  & 1.19 & 1.19 & 1.00 & any  & 1   & 0.33$^a$ \\ \hline
4  & 0.99 & 1.28 & 1.29 & 1.13 & 32  & 2.40     \\ \hline
6  & 0.90 & 1.44 & 1.60 & 1.13 & 56  & 42.6     \\ \hline
8  & 0.81 & 1.61 & 1.99 & 1.13 & 76  & 836      \\ \hline
10 & 0.80 & 1.74 & 2.18 & 1.13 & 104 & 13,150
\end{tabular}
\vspace{0.5cm}
\caption{ Parameters of the I-V characteristics shown in Fig.~\ref{fig7}. 
$R$ is the resistance of one molecule at $V = +V_{{\rm op}}$. 
\newline
$^a$ resistance at $V = 1.6$V. }
\label{table3}
\end{table}

The important issue is the exponential growth of the diode resistance with 
$n $, see the last column of Table~\ref{table3}. This result is, of course,
expected because the transmission as well as the current through the
molecule is directly proportional to $\Gamma_R$, see Eqs.~(\ref{eq:TE}) 
and (\ref{eq:Ires}) at $\Gamma_L \gg \Gamma_R$. As the barrier length 
increases, the probability of underbarrier tunneling goes down 
exponentially fast, which is reflected in the numerical data.  According 
to our calculations, addition of every extra pair of -(CH$_2$)- groups 
increases the resistance by a factor of $17 \pm 2$. 
[Note that for $(n=2) \rightarrow (n=4)$ the transition should not 
follow the same trend because at $(n=2)$ the molecule is symmetric and 
$\Gamma_L = \Gamma_R$.] Thus we have an unfavorable trade off problem where 
the rectification ratio improves linearly with $n$ but the resistance and 
the time constant of the device worsens exponentially. This is an inherent 
feature of our mechanism and in some sense the price for its simplicity. 
When discussing the rectifier resistances, one should not forget that what 
matters is the total resistance of an electronic element. With a typical 
target size of $(10 \times 10)$ nm$^2$, the number of molecules per 
junction is going to be of order of 1000. Then even $n=8$ and 10 
diodes will have total resistances of 1-10 M$\Omega$, which amounts to 
a time constant $\sim 10^{-10} s$ for a $1 \mu$m long nanowire, and
$\sim 10^{-6} s$ for a 1 cm long nanowire.

\begin{table}[t]
\begin{tabular}{|c||c|c|c|c|c|c|}
$W$ (eV) & 4.0  & 4.2 & 4.5 & 5.0 & 5.2 & 5.5            \\ \hline
$\Delta$ (eV) & 0.55 & 0.75 & 1.05 & 1.55 & 1.75 & 2.05 \\ \hline\hline
(1,2)   & 4$^a$ & 5$^b$ & 8$^c$ & 12$^d$ & 39 & 56 \\ \hline
(1,4) & 10$^a$ & 13$^b$ & 26$^c$ & 75 & 111 & 154 \\ \hline
(1,6) & 15 & 25 & 44 & 112 & 127 & 152 \\ \hline
(1,8) & 16 & 25 & 44 & 91 & 143$^e$ & 126 \\ \hline
(1,10) & 19 & 27 & 47 & 85 &  & 123 \\ \hline\hline
(2,4) & 32 & 52 & 91 & 138 & 156 & 154 \\ \hline
(2,6) & 56 & 74 & 173 & 276 & 374 & 444 \\ \hline
(2,8) & 76 & 92 & 225 & N/C & 438$^e$ & N/C \\ \hline
(2,10) & 104 & 118 & 290 & N/C & {\bf 556}$^e$ & N/C \\ \hline\hline
$V_{{\rm op}}$ (V) & 1.13 & 1.51 & 2.08 & 3.02 & 3.50 & 3.97
\end{tabular}
\vspace{0.5cm}
\caption{ The rectification ratio $I_{+}/I_{-}$ for six values of the wire
work function and nine different molecules. The first column is the $(m,n)$
numbers of the diodes. The top row is the work function of the electrodes
in eV. The bottom row is the operating voltage $V_{{\rm op}}$ in volts, at
which the rectification ratio is reported. ``N/C'' stands for non-conclusive
evidence due to numerical noise in calculating small currents under reverse
bias. Best rectification is achieved for the (2,10) MR at $W = 5.2$ eV 
(in bold). The $W$ value is close to the work function of gold. 
\newline
$^a$ at $V = 1.32$ V. \newline
$^b$ at $V = 1.89$ V. \newline
$^c$ at $V = 2.46$ V. \newline
$^d$ at $V = 3.59$ V. \newline
$^e$ at $V = 3.02$ V. }
\label{table4}
\end{table}

The data from Table~\ref{table2} suggests that rectification could be
improved further by increasing the initial energy difference $\Delta$. 
We computed I-V characteristics for every value of the work function 
listed in Table~\ref{table2} and for all MRs $(m,n)$ with $m=1,2$ and 
$n= 2, 4, 6, 8, 10$. [In addition, everything was calculated for the 
two values of the electrode matrix element $W_{ss\sigma}$, bringing the 
total number of studied I-Vs to 120.]  All the curves have shapes similar to 
those in Figs.~\ref{fig6} and \ref{fig7}, therefore we do not show 
them explicitly. Table~\ref{table4} summarizes the results on 
rectification ratios for a variety of combinations.  As expected,
the rectification ratio grows with both $W$ and $n$.  The maximum
ratio of above 500 was observed for $n=10$ and $W = 5.2$ eV 
($\Delta = 1.75$ eV).

\begin{figure}[t]
\begin{center}
\leavevmode
\hbox{
\epsfxsize=8.6cm
\epsffile{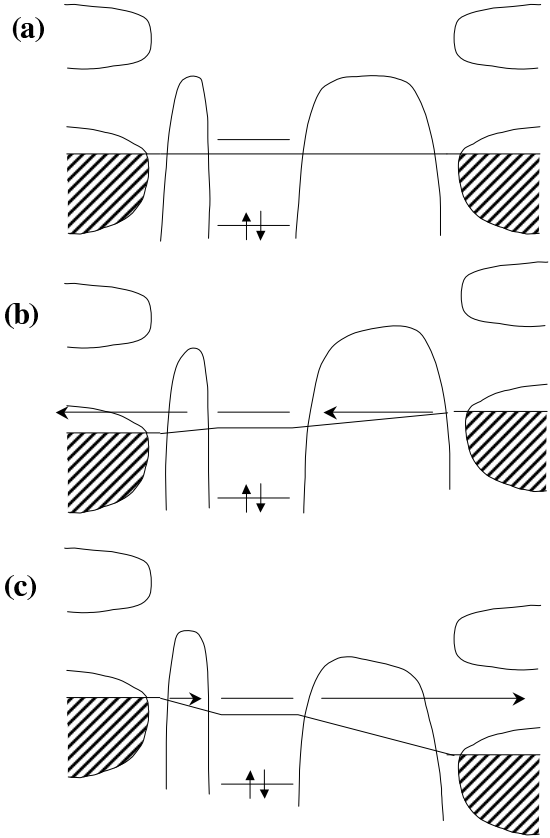}
}
\end{center}
\vspace{0.0cm}
\caption{ (a) A gap in the density of states of the electrodes. 
(b) Under forward bias, when the right Fermi level aligns with the LUMO, 
there are still states available in the left electrode. The current flows. 
(c) Under reverse bias, by the time the left Fermi level aligns with the 
LUMO, there are no states available in the right electrode. The elastic 
current is blocked. }
\label{fig8}
\end{figure}

{\em (iv) Role of energy gaps in the electrodes' density of states.} This
effect is not generic and potentially applies to semiconductor electrodes
only. Nevertheless, it offers an interesting possibility to enhance the
rectification property of the diode by inhibiting current flow in the reverse
direction due to an energy gap in the density of states. The basic idea is
illustrated in Fig.~\ref{fig8}. At some voltage $V_g$, the bottom of the gap
aligns with the conducting molecular orbital. At higher $V > V_g$ resonant
tunneling becomes impossible due to the lack of available final electronic
states. Accordingly, the current must drop to zero at $V = V_g$. If $V_g$
lies in between $V_F$ and $V_R$, then there will be some substantial current
in the forward direction in the interval $V_F < V < V_g$, Fig.~\ref{fig8}(b),
while no current at all in the reverse direction, Fig.~\ref{fig8}(c). In the
latter case, the gap of the right electrode is reached {\em before} the left
Fermi energy aligns with the conducting orbital.

\begin{figure}[t]
\begin{center}
\leavevmode
\hbox{
\epsfxsize=8.6cm
\epsffile{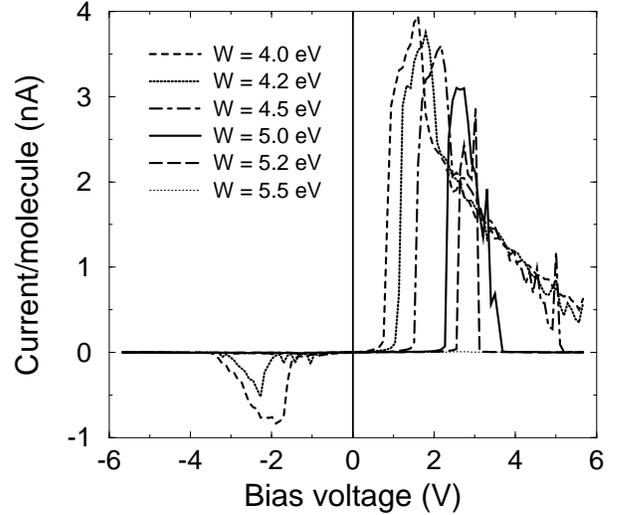}
}
\end{center}
\vspace{0.0cm}
\caption{I-V characteristic of the (2,8) MR for the electrode
matrix element $W_{ss\protect\sigma} = 0.909$ eV. A gap in the 
electrode's density of states cuts off current at large voltages.
Apart from current rectification, this leads to a significant 
negative differential resistance effect.  The latter will be reduced
if inelastic processes are taken into account. }
\label{fig9}
\end{figure}

The above argument does not take into account inelastic processes. If
included, they would result in a non-zero current even in the situation of
Fig.~\ref{fig8}(c). In this case, the electrons can tunnel under the right
barrier with irradiation of phonons, arriving at the right electrode with a
reduced energy for which there are available states at the top of the
valence band. In this paper we assume that the elastic tunneling is the
dominant transport mechanism and such processes can be neglected. 
We have modeled this effect by using the smaller electrode matrix element 
$W_{ss\sigma} = 0.909$ eV.  Typical I-V characteristics are shown
in Fig.\ref{fig9} and their parameters are summarized in Table~\ref{table5}.
With increasing $W$ (and $\Delta$), the forward and reverse voltages 
$V_F$ and $V_R$ increase in accordance with Eqs.~(\ref{two}) and 
(\ref{three}).  At the same time, $V_g$ {\em decreases} because the
bottom of the energy gap moves closer to the conducting molecular level.
As a result, the window of resonant tunneling is squeezed from both
sides and at some critical $W$ vanishes altogether.  This effect can be
seen in Fig.\ref{fig9}.  In the reverse direction, the peak disappears
between $W = 4.2$ eV and $W = 4.5$ eV.  In the forward direction,
the same happens between $W = 5.2$ eV and $W = 5.5$ eV. (Of course,
resonant tunneling survives longer in the forward direction because
$V_F > V_R$.)   

\begin{table}[b]
\begin{tabular}{|c||c|c|c|c|c|c|}
$W$ (eV) & $\Delta$ (eV) & $V_F$ (V) & $V_R$ (V) & $V_g$ (V) & $V_{\rm
op}$ (V)
& $I_{+}/I_{-}$ \\ \hline\hline
4.0 & 0.55 & 0.84 & 1.62 & N/C  & 1.23 &   70 \\ \hline
4.2 & 0.75 & 1.17 & 2.16 & N/C  & 1.80 &  151 \\ \hline
4.5 & 1.05 & 1.58 & N/C  & 5.20 & 2.17 &  282 \\ \hline
5.0 & 1.55 & 2.33 & N/C  & 3.68 & 2.83 & 1308 \\ \hline
5.2 & 1.75 & 2.60 & N/C  & 3.10 & 2.83 & 1064 \\ \hline
5.5 & 2.05 & N/C  & N/C  & N/C  & 2.83 &    3
\end{tabular}
\vspace{0.5cm}
\caption{ 
Parameters of the I-V characteristics of the (2,8) molecular rectifier
for $W_{ss\protect\sigma} = 0.909$ eV. (I-Vs are shown in Fig.~\ref{fig9}.) 
Notice how the rectification ratio exceeds 1000 when the resonant
conduction disappears in the reverse direction, but then drops
to just 3 when it disappears in the forward direction too.
}
\label{table5}
\end{table}

Notice that the I-Vs of Fig.~\ref{fig9} display large negative
differential resistance under positive bias. In this respect, our molecular
diodes behave similarly to the conventional semiconducting tunneling diodes.
However, under negative bias, the behavior is quite different. While the
tunneling diodes conduct very well under reverse bias, our MR conduct 
very little.

\section{Conclusions}
\label{sec:five}

The major purpose of this paper has been to predict several {\em trends} 
in the I-V characteristics of MR when they are dominated by 
resonant tunneling through a certain molecular orbital localized on the
conjugated part of a spatially asymmetric molecule. To achieve a
large rectification ratio, the conjugated part of the molecule must be
connected to electrodes by insulating molecular groups of different length.
By varying the ratio between the barrier lengths, one can achieve a
rectification ratio of several hundred while keeping the current through 
the molecule at measurable levels.  

As far as the described mechanism of molecular rectification is concerned,
the observed trends could be divided into ``spatial'' and ``energetic'' 
domains.  The spatial trends refer to the changes in the I-V
characteristic that follow from changing the lengths of the two
insulating barriers of the MR.  The length of the shorter barrier
controls the width the molecular resonance.  We have found the optimal
length to be two -(CH$_2$)- groups.  One such group does not provide
sufficient insulation of the conducting unit from the metal.  As a result,
the resonance is too broad and current asymmetry is not pronounced.
More than two -(CH$_2$)- groups make the resonance narrower than 
temperature and disorder induced widths, which is not useful.
Increasing the length of the {\em longer} barrier improves the current
rectification ratio but increases the total resistance of the molecular
diode.  These trends can be observed experimentally by studying 
several molecules with different lengths of the longer chain but
identical otherwise.  The energetic trends refer to the dependence 
of the rectification property on the work function of the electrode
material and on the electron affinity of the molecule.  The two 
quantities define the parameter $\Delta$, which is the most important
energy scale in the problem.  In general, larger $\Delta$ leads to
better rectification but at the same time to larger operating 
voltages.  These trends can also be checked experimentally by
measuring, for instance, the same molecule on gold and silver
electrodes, or by measuring the molecules with different conducting
units (say, benzene and naphthalene) on the same set of electrodes.

\begin{figure}[t]
\begin{center}
\leavevmode
\hbox{
\epsfxsize=8.6cm
\epsffile{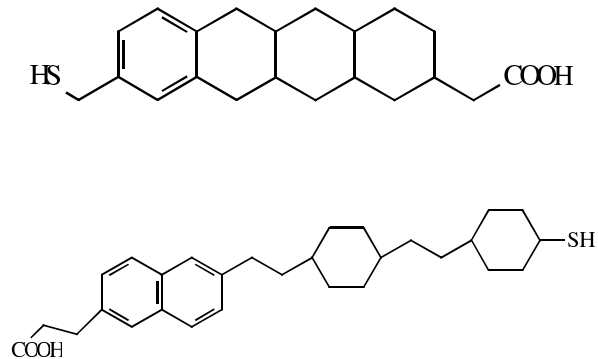}
}
\end{center}
\vspace{-0.5cm}
\caption{ Other possible molecular diodes. }
\label{fig10}
\end{figure}

Most of our results are rather insensitive to changing the end (``anchor'') 
groups. The role of the latter is to provide better connection to the 
electrodes. Thus the end groups could be adjusted to the particular 
experimental setup without significantly altering the electrical properties 
of the device. They could even be absent altogether if the measurement
technique does not require covalent bonding between the molecules and the
electrodes. The composition of the insulating barriers may be changed too.
For instance, instead of alkane chains one could use a combination of the
alkane segments and saturated cyclic hydrocarbons, see Fig.~\ref{fig10}. 
An advantage of this design is that the molecules have more or less the same
cross section along their entire length. Therefore the packing of the film
will be much better. On the other hand, such MR could be harder to
synthesize.

These conclusions are certainly limited to the particular set of
approximations used in the present study. For instance, we have ignored
inelastic processes, which should become progressively more important for
larger molecules and at larger bias voltages.  Inelastic scattering
generally increases the current through the system,
since the carrier has more final states to tunnel into when it can
lose energy by exciting a vibronic excitation. In the present resonant
case, when $V$ exceeds the threshold, either $V_{F}$ or $V_{R}$, there will
be some electrons with energy {\em higher} than the energy of the LUMO
resonance. Such electrons can still tunnel through the junction effectively
by first loosing their energy and then tunneling resonantly. Thus the
overall current will {\em continue to grow} after the initial sharp
increase, in contrast to the elastic-only case where the current stays
approximately constant. Additionally, a finite temperature should smoothen
the sharp features of the I-V characteristic, thereby reducing the 
rectification ratio. The inclusion of inelastic processes would probably
reduce the estimated rectification ratio. 

In conclusion, we have described a simple principle for molecular 
rectification.  It requires only one conducting molecular level located 
asymmetrically with respect to the electrodes.  Theoretical calculations 
have shown that current rectification ratios of several hundred are 
achievable with this mechanism.  Higher ratios are likely to be 
accompanied by a significant reduction in conductance, which will make 
such diodes less practical.  We have made several predictions of how the 
rectification properties should change between different molecules and 
different electrode materials (work functions).  These trends are 
experimentally verifiable on families of structurally similar molecules.  
Such trends are signatures of resonant tunneling through molecules 
and could serve as proof of experimental observation of the latter.

\end{document}